\documentclass[preprintnumbers,superscriptaddress,showpacs,pra]{revtex4}
\usepackage{amssymb,amsmath}
\usepackage{graphicx}
\usepackage{amsmath}
\usepackage{amsfonts}

\input{tcilatex}
\begin{document}

\preprint{}
\title[]{Posmon spectrosopy of quantum state on a circle}
\author{Q. H. Liu}
\email{quanhuiliu@gmail.com}
\affiliation{School for Theoretical Physics, and Department of Applied Physics, Hunan
University, Changsha, 410082, China}
\date{\today }

\begin{abstract}
Developing the analysis of the distribution of the particle's
position-momentum dot product, the so-called \textit{posmom} $\mathbf{x}%
\cdot \mathbf{p}$\textbf{,} to quantum states on a circular circle on
two-dimensional Cartesian coordinates, we give its posmometry (introduced
recently by Y. A. Bernard and P. M. W. Gill, Posmom: The Unobserved
Observable, J. Phys. Chem. Lett. 1\textbf{(}2010\textbf{)}1254) for
eigenstates of the free motion on the circle, i.e., $z$-axis component of
the angular momentum. The posmom has two parity symmetries, specifically,
invariant under two operations $m_{x}$ and $m_{y}$ representing mirror
symmetry about $x$ and $y$ axis respectively. The complete eigenfunction set
of the posmom is then four-valued and consists of four basic parts each of
them is defined within a distinct quadrant of the circle. The results are
not only potentially experimentally testable, but also reflect a fact that
the embedding of the circle $S^{1}$ in two-dimensional flat space $R^{2}$ is
physically reasonable.
\end{abstract}

\maketitle

\section{Introduction}

Recently, Gill et al. introduce a new operator, the particle's
position-momentum dot product $\mathbf{x}\cdot\mathbf{p}$, or \textit{posmom}
as they called, and establish a \textit{posmometry} (the distribution
density of the posmom) for some atomic and molecular systems. \cite%
{posmometry1,posmometry2} The \textit{posmom }operator in one of the
Cartesian axes, say, $x$ axis $Q_{x}\equiv(xp_{x}+p_{x}x)/2$, is an
essentially self-adjoint operator. \cite{winter,milburn} In Ref.[%
\onlinecite{liu141}], we developed this operator on the a two-dimensional
spherical surface $S^{2}$ and successfully worked out its distribution
densities for some molecular rotational states. In this Letter, we explore
the posmometry of quantum states on a circle which frequently model the
planar rigid rotor, molecular rotation constrained on a plane, etc..

As embedding $S^{1}$ in the two-dimensional flat space $R^{2}$, there are
two operators $Q_{i}$ $(i=x,y)$ that are respectively defined along two
Cartesian axes of coordinate respectively, which turn out to take following
form, 
\begin{equation}
Q_{i}=\frac{1}{2}\left( x_{i}p_{i}+p_{i}x_{i}\right) ,
\end{equation}
where $\mathbf{x}=(r\cos\varphi,r\sin\varphi),\varphi\in(0,2\pi)$, and $%
p_{x}=i\hbar/r\left( \sin\varphi\partial_{\varphi}+\cos\varphi/2\right) $, $%
p_{y}=-i\hbar/r\left( \cos\varphi\partial_{\varphi}-\sin\varphi/2\right) $.
In fact, the momenta $p_{x}$ and $p_{y}$ are special case of the the
so-called geometric momentum $\mathbf{p}=-i\hbar(\nabla_{surf}+M\mathbf{n/}2)
$ \cite{liu133} on an $N$-dimensional surface which is embedded in $N+1$
dimensional Euclidean space, where $\nabla_{surf}$ is the gradient operator
on surface, and $M$ is the mean curvature and $\mathbf{n}$ is the normal
vector. \cite{liu133} The explicit forms of $Q_{x}$ and $Q_{y}$ are,%
\begin{equation}
Q_{x}=\frac{i\hbar}{2}(\sin2\varphi\frac{\partial}{\partial\varphi}%
+\cos2\varphi),\text{ }Q_{y}=-Q_{x}.   \label{indep}
\end{equation}

In the following section II, I\ will present the elementary properties of
this operator. In section III, I will give the posmometry for eigenstates of
the $z$-component angular momentum $L_{z}:\Phi_{m}\left( \varphi\right)
=\exp(im\varphi)/\sqrt{2\pi}$, $m=0,\pm1,\pm2,...$. Final section VI is our
conclusions.

\section{Elementary properties of the posmom operator}

The following properties of the two posmom operators $Q_{i}$ $(i=1,2)$ are
easily attainable.

i) Since the geometric momentum $\mathbf{p}$ describes the motion
constrained on the surface $S^{1}$ and there is no motion along the normal
direction $\mathbf{n}$, which in quantum mechanics is expressed by $\mathbf{%
x\cdot p+p\cdot x}\equiv2(Q_{x}+Q_{y})=0$ while it is in classical mechanics
expressed by $\mathbf{x\cdot p}=0$. This is why two operators $Q_{x}$ and $%
Q_{y}$ are linearly dependent, as shown in Eq. (\ref{indep}). So it suffices
to study one of the them, and I\ will concentrate $Q_{x}$.

ii) The operator $Q_{x}$ has the reflection symmetry: 
\begin{equation}
Q_{x}(\Theta\pm\varphi)=Q_{x}(\Theta+\varphi),\text{ }\Theta=0,\frac{\pi}{2}%
,\pi,\frac{3\pi}{2}.
\end{equation}
In other words, posmom $Q_{x}$ commutes with two parity operators $m_{x}$
and $m_{y}$ denoting two reflections about axes $x=0$, $y=0$ respectively,
and we have, 
\begin{equation}
\left[ m_{x},Q_{x}\right] =\left[ m_{y},Q_{x}\right] =0.
\end{equation}
This mirror invariance is helpful in construction of the complete set of the
eigenfunctions on the circle once the eigenfunctions of $Q_{x}$ in four
quadrants\ $\varphi\in(0,\pi/2),$ $(\pi/2,\pi)$, $(\pi,3\pi/2)$ and $%
(3\pi/2,2\pi)$ are known, respectively.

iii) In the full circle\ $\varphi \in (0,2\pi )$, we have the solution to
the eigenvalue problem $Q_{x}\xi _{\lambda }(\varphi )=\hbar \lambda \xi
_{\lambda }(\varphi )$, 
\begin{equation}
\xi _{\lambda }(\varphi )=\frac{1}{\sqrt{\pi }}\frac{1}{\sqrt{\left\vert
\sin 2\varphi \right\vert }}\exp (-i\lambda \ln \left\vert \tan \varphi
\right\vert )\text{, }\lambda \in (-\infty ,\infty ),
\end{equation}%
which is delta function normalized in any one of four quadrants, say in the
first: $\int_{0}^{\pi /2}\xi _{\lambda ^{\prime }}^{\ast }(\varphi )\xi
_{\lambda }(\varphi )d\varphi =\delta (\lambda ^{\prime }-\lambda )$. For
convenience, we can use $\left\vert \xi _{\lambda }^{J}\right\rangle $ ($%
J=I,II,III,IV$) to denote the eigenfunctions defined within the $J$th
quadrants of the circle respectively.

iv) Note that $\left\vert \xi_{\lambda}^{J}\right\rangle $ is not the
simultaneous eigenfunction of operators $Q_{x}(\varphi)$, $m_{x}$ and $m_{y}$%
. The complete simultaneous eigenfunction set of operators $Q_{x}(\varphi)$, 
$m_{x}$ and $m_{y}$ is given by,
\begin{subequations}
\begin{align}
\left\vert \psi_{\lambda}^{xy}\right\rangle & =\left( \left\vert
\xi_{\lambda}^{I}\right\rangle +\left\vert \xi_{\lambda}^{II}\right\rangle
+\left\vert \xi_{\lambda}^{III}\right\rangle +\left\vert
\xi_{\lambda}^{VI}\right\rangle \right) /2,  \label{parity1} \\
\left\vert \psi_{\lambda}^{\overline{x}\overline{y}}\right\rangle & =\left(
\left\vert \xi_{\lambda}^{I}\right\rangle -\left\vert
\xi_{\lambda}^{II}\right\rangle +\left\vert \xi_{\lambda}^{III}\right\rangle
-\left\vert \xi_{\lambda}^{VI}\right\rangle \right) /2,  \label{parity2} \\
\left\vert \psi_{\lambda}^{\overline{x}y}\right\rangle & =\left( \left\vert
\xi_{\lambda}^{I}\right\rangle +\left\vert \xi_{\lambda}^{II}\right\rangle
-\left\vert \xi_{\lambda}^{III}\right\rangle -\left\vert
\xi_{\lambda}^{VI}\right\rangle \right) /2,  \label{parity3} \\
\left\vert \psi_{\lambda}^{x\overline{y}}\right\rangle & =\left( \left\vert
\xi_{\lambda}^{I}\right\rangle -\left\vert \xi_{\lambda}^{II}\right\rangle
-\left\vert \xi_{\lambda}^{III}\right\rangle +\left\vert
\xi_{\lambda}^{VI}\right\rangle \right) /2,   \label{parity4}
\end{align}
where $x$ and $\overline{x}$ indicate even and odd parity respectively about 
$x$-axis, and so on, 
\end{subequations}
\begin{subequations}
\begin{align}
m_{x}m_{y}\left\vert \psi_{\lambda}^{xy}\right\rangle & =m_{y}\left\vert
\psi_{\lambda}^{xy}\right\rangle =\left\vert
\psi_{\lambda}^{xy}\right\rangle ,m_{x}m_{y}\left\vert \psi_{\lambda}^{%
\overline{x}\overline{y}}\right\rangle =-m_{y}\left\vert \psi_{\lambda}^{%
\overline{x}\overline{y}}\right\rangle =\left\vert \psi_{\lambda}^{\overline{%
x}\overline{y}}\right\rangle , \\
m_{x}m_{y}\left\vert \psi_{\lambda}^{\overline{x}y}\right\rangle &
=m_{x}\left\vert \psi_{\lambda}^{\overline{x}y}\right\rangle =-\left\vert
\psi_{\lambda}^{\overline{x}y}\right\rangle ,m_{x}m_{y}\left\vert
\psi_{\lambda}^{x\overline{y}}\right\rangle =m_{y}\left\vert \psi_{\lambda
}^{x\overline{y}}\right\rangle =-\left\vert \psi_{\lambda}^{x\overline{y}%
}\right\rangle .
\end{align}
The orthonormality and completeness of the eigenfunction set (\ref{parity1}%
)-(\ref{parity4}) are satisfied. I.e., we have following two relations, 
\end{subequations}
\begin{equation}
1,\text{ }\left\langle
\psi_{\lambda^{\prime}}^{i^{\prime}j^{\prime}}\left\vert
\psi_{\lambda}^{ij}\right\rangle \right. =\delta(\lambda^{\prime
}-\lambda)\delta_{i^{\prime}i}\delta_{j^{\prime}j},
\end{equation}
where $(i,j)$ standing for $x,y,\overline{x}$, and $\overline{y}$. For any
state $\Phi\left( \varphi\right) $ on the circle, we have, 
\begin{equation}
2,\text{ }\Phi\left( \varphi\right) =\int_{-\infty}^{\infty}\alpha\left(
\lambda\right) \psi_{\lambda}^{xy}\left( \varphi\right) d\lambda
+\int_{-\infty}^{\infty}\beta\left( \lambda\right) \psi_{\lambda}^{\overline{%
x}\overline{y}}\left( \varphi\right) d\lambda+\int_{-\infty
}^{\infty}\mu\left( \lambda\right) \psi_{\lambda}^{\overline{x}y}\left(
\varphi\right) d\lambda+\int_{-\infty}^{\infty}\upsilon\left( \lambda
\right) \psi_{\lambda}^{x\overline{y}}\left( \varphi\right) d\lambda , 
\label{complete2}
\end{equation}
where, $\int\left( \left\vert \alpha\left( \lambda\right) \right\vert
^{2}+\left\vert \beta\left( \lambda\right) \right\vert ^{2}+\left\vert
\mu\left( \lambda\right) \right\vert ^{2}+\left\vert \upsilon\left(
\lambda\right) \right\vert ^{2}\right) d\lambda=1$. This normalization
clearly states that for a given $\lambda$, the distribution density $%
p(\lambda)$ usually comes form four parts, 
\begin{equation}
p(\lambda)=\left\vert \alpha\left( \lambda\right) \right\vert
^{2}+\left\vert \beta\left( \lambda\right) \right\vert ^{2}+\left\vert
\mu\left( \lambda\right) \right\vert ^{2}+\left\vert \upsilon\left(
\lambda\right) \right\vert ^{2}\text{.}   \label{p}
\end{equation}
It means that for a given $\lambda$, the probability amplitude is from Eq. (%
\ref{complete2}) a four-valued function, 
\begin{subequations}
\begin{align}
\alpha\left( \lambda\right) & =\int_{0}^{2\pi}\psi_{\lambda}^{xy\ast }\left(
\varphi\right) \Phi\left( \varphi\right) d\varphi,\text{ }\beta\left(
\lambda\right) =\int_{0}^{\pi/2}\psi_{\lambda}^{\overline {x}\overline{y}%
\ast}\left( \varphi\right) \Phi\left( \varphi\right) d\varphi, \\
\mu\left( \lambda\right) & =\int_{0}^{2\pi}\psi_{\lambda}^{\overline {x}%
y\ast}\left( \varphi\right) \Phi\left( \varphi\right) d\varphi,\text{ }%
\upsilon\left( \lambda\right) =\int_{0}^{\pi/2}\psi_{\lambda}^{x\overline{y}%
\ast}\left( \varphi\right) \Phi\left( \varphi\right) d\varphi.
\end{align}
However, in the following section, we see that for the eigenstates $\Phi
_{m}\left( \varphi\right) =\exp(im\varphi)/\sqrt{2\pi}$ of the $z$-component
angular momentum $L_{z}$, the probability amplitude is in general
triple-valued.

\section{Posmometry for eigenstates of the $z$-component angular momentum $%
L_{z}$}

As is well known, the $z$-component angular momentum $L_{z}=-i\hbar \partial
_{\varphi }$ has a complete set of eigenfunctions $\Phi _{m}\left( \varphi
\right) =\exp (im\varphi )/\sqrt{2\pi }$ ($m=0$, $\pm 1$, $\pm 2$, ...) that
span a Hilbert space for analyzing any state on $S^{1}$. In general, we
have, 
\end{subequations}
\begin{equation*}
\frac{1}{\sqrt{2\pi }}\exp (im\varphi )=\int_{0}^{2\pi }\alpha _{m}\left(
\lambda \right) \psi _{\lambda }^{xy}d\lambda +\int_{0}^{2\pi }\beta
_{m}\left( \lambda \right) \psi _{\lambda }^{\overline{x}\overline{y}%
}d\lambda +\int_{0}^{2\pi }\mu _{m}\left( \lambda \right) \psi _{\lambda }^{%
\overline{x}y}d\lambda +\int_{0}^{2\pi }\upsilon _{m}\left( \lambda \right)
\psi _{\lambda }^{x\overline{y}}d\lambda 
\end{equation*}%
where the expansion coefficients $\alpha _{m}\left( \lambda \right) $, $%
\beta _{m}\left( \lambda \right) $, $\mu _{m}\left( \lambda \right) $ and $%
\upsilon _{m}\left( \lambda \right) $ are given by, 
\begin{subequations}
\begin{align}
\alpha _{m}\left( \lambda \right) & =\frac{1+(-1)^{m}}{2\pi \sqrt{2}}\left(
I_{m}\left( \lambda \right) +\exp \left( \frac{im\pi }{2}\right) I_{m}\left(
-\lambda \right) \right) , \\
\beta _{m}\left( \lambda \right) & =\frac{1+(-1)^{m}}{2\pi \sqrt{2}}\left(
I_{m}\left( \lambda \right) -\exp \left( \frac{im\pi }{2}\right) I_{m}\left(
-\lambda \right) \right) , \\
\mu _{m}\left( \lambda \right) & =\frac{1-(-1)^{m}}{2\pi \sqrt{2}}\left(
I_{m}\left( \lambda \right) +\exp \left( \frac{im\pi }{2}\right) I_{m}\left(
-\lambda \right) \right) , \\
\upsilon _{m}\left( \lambda \right) & =\frac{1-(-1)^{m}}{2\pi \sqrt{2}}%
\left( I_{m}\left( \lambda \right) -\exp \left( \frac{im\pi }{2}\right)
I_{m}\left( -\lambda \right) \right) ,
\end{align}%
with,
\end{subequations}
\begin{align}
I_{m}\left( \lambda \right) & =\int_{0}^{\pi /2}\frac{\exp (im\varphi )}{%
\sqrt{\sin 2\varphi }}\exp (i\lambda \ln \tan \varphi )d\varphi  \\
=& \left( \frac{1}{2}+\frac{i}{2}\right) e^{-\frac{\pi \lambda }{2}}\Gamma
\left( \frac{m+1}{2}\right) \left\{ f(\lambda ,m)-i^{m+1}f^{\ast }(\lambda
,m)\right\} ,
\end{align}%
in which with $F(a,b;c;z)$ symbolizing the hypergeometric function, 
\begin{equation}
f(\lambda ,m)=\frac{\Gamma \left( \frac{1}{2}+i\lambda \right) }{\Gamma
\left( \frac{m}{2}+1+i\lambda \right) }\,F\left( \frac{1}{2}+i\lambda ,\frac{%
m+1}{2};\frac{m}{2}+1+i\lambda ;-1\right) .
\end{equation}%
Because of the eigenfunctions $\Phi _{m}\left( \varphi \right) $ can be
decomposed into two parts according to mirror symmetry operators: $\exp
(im\varphi )/\sqrt{2\pi }=\left\{ \cos (m\varphi )+i\sin (m\varphi )\right\}
/\sqrt{2\pi }$, it is for our purpose sufficient to study the eigenfunctions 
$\Phi _{m}\left( \varphi \right) $ with $m\succeq 0$. Evidently, for $m$
being a positive even number $m=2k$ ($k=1,2,3,...$), we obtain, 
\begin{subequations}
\begin{align}
\alpha _{2k}\left( \lambda \right) & =\frac{1}{\sqrt{2}\pi }\left(
I_{2k}\left( \lambda \right) +\left( -1\right) ^{k}I_{2k}\left( -\lambda
\right) \right) , \\
\beta _{2k}\left( \lambda \right) & =\frac{1}{\sqrt{2}\pi }\left(
I_{2k}\left( \lambda \right) -\left( -1\right) ^{k}I_{2k}\left( -\lambda
\right) \right) , \\
\mu _{2k}\left( \lambda \right) & =\upsilon _{2k}\left( \lambda \right) =0.
\end{align}%
For $m$ being a positive odd number $m=2k+1$ ($k=1,2,3,...$), we obtain, 
\end{subequations}
\begin{subequations}
\begin{align}
\mu _{2k+1}\left( \lambda \right) & =\frac{1}{\sqrt{2}\pi }\left(
I_{2k+1}\left( \lambda \right) +i\left( -1\right) ^{k}I_{2k+1}\left(
-\lambda \right) \right) , \\
\upsilon _{2k+1}\left( \lambda \right) & =\frac{1}{\sqrt{2}\pi }\left(
I_{2k+1}\left( \lambda \right) -i\left( -1\right) ^{k}I_{2k+1}\left(
-\lambda \right) \right) , \\
\alpha _{2k+1}\left( \lambda \right) & =\beta _{2k+1}\left( \lambda \right)
=0.
\end{align}%
So, we see that the probability amplitude is in general triple-valued.
Unfortunately, the expansion coefficients ($\alpha $, $\beta $, $\mu $, $%
\upsilon $), provided nontrivial, can not be all greatly simplified unless $%
m=0$ and $m$ being odd. For $m=0$ and $m=2k$, we have following relations, 
\end{subequations}
\begin{subequations}
\begin{align}
\alpha _{0}\left( \lambda \right) & =\frac{\left\vert \Gamma \left(
1/4-i\lambda /2\right) \right\vert ^{2}}{2\pi ^{3/2}},\beta _{0}\left(
\lambda \right) =0, \\
\alpha _{2k}\left( \lambda \right) & \neq \beta _{2k}\left( \lambda \right) ,
\end{align}%
where $m=0$ is the only case the probability amplitude of the posmom is
double valued. For $m$ being odd we have $\left\vert \mu _{2k+1}\left(
\lambda \right) \right\vert ^{2}=\left\vert \upsilon _{2k+1}\left( \lambda
\right) \right\vert ^{2}$, and for $m=1,3,5$, we have explicitly,
\end{subequations}
\begin{subequations}
\begin{align}
\mu _{1}\left( \lambda \right) & =\frac{i}{\sqrt{2}\left( \cosh (\pi \lambda
/2)-i\sinh (\pi \lambda /2)\right) },\text{ }\upsilon _{1}\left( \lambda
\right) =-i\text{ }\mu _{1}^{\ast }\left( \lambda \right) ,\text{ }%
p_{1}(\lambda )=\sec h(\lambda \pi ), \\
\mu _{3}\left( \lambda \right) & =\frac{\sqrt{2}\lambda }{\cosh \left( \pi
\lambda /2\right) -i\sinh \left( \pi \lambda /2\right) },\text{ }\upsilon
_{3}\left( \lambda \right) =-i\text{ }\mu _{3}^{\ast }\left( \lambda \right)
,\text{ }p_{3}(\lambda )=4\lambda ^{2}\sec h(\lambda \pi ), \\
\mu _{5}\left( \lambda \right) & =\frac{-(1-4\lambda ^{2})i}{2\sqrt{2}\left(
\cosh \left( \pi \lambda /2\right) -i\sinh \left( \pi \lambda /2\right)
\right) },\text{ }\upsilon _{5}\left( \lambda \right) =-i\text{ }\mu
_{5}^{\ast }\left( \lambda \right) ,\text{ }p_{5}(\lambda )=\frac{1}{4}%
\left( 1-4\lambda ^{2}\right) ^{2}\sec h(\lambda \pi ).
\end{align}

The probability distributions $p(\lambda )$ (\ref{p}) for rotational states
represented by $e^{im\varphi }/\sqrt{2\pi }$ with $m=0$ to $6,40,41$ are
plotted in Fig. 1 ($m=0,2$), Fig. 2 ($m=4$), Fig. 3 ($m=6$), Fig. 4 ($m=1,3,5
$), Fig. 6 ($m=40$) and Fig. 6 ($m=41$) respectively. On the whole, they are
similar to the momentum distributions of stationary states for the
one-dimensional simple harmonic oscillator. It is understandable from an
examination of the free motion on the circle. In classical mechanics for the
free motion with a frequency $\omega $, we have $x=r\cos (\omega t)$ and $%
p_{x}=-m\omega r\sin (\omega t)$ and therefore $xp_{x}=-(m\omega
r^{2}/2)\sin (2\omega t)$. Then, in a classical state, the posmom $xp_{x}$
has a half period as $x$ or $p_{x}$ has. In classical limit, whenever $m$
being even or odd number, the eigenstate $e^{im\varphi }/\sqrt{2\pi }$
behaves like a simple harmonic oscillator as seen from the posmom.


\begin{figure}
\includegraphics[width=0.9\textwidth]{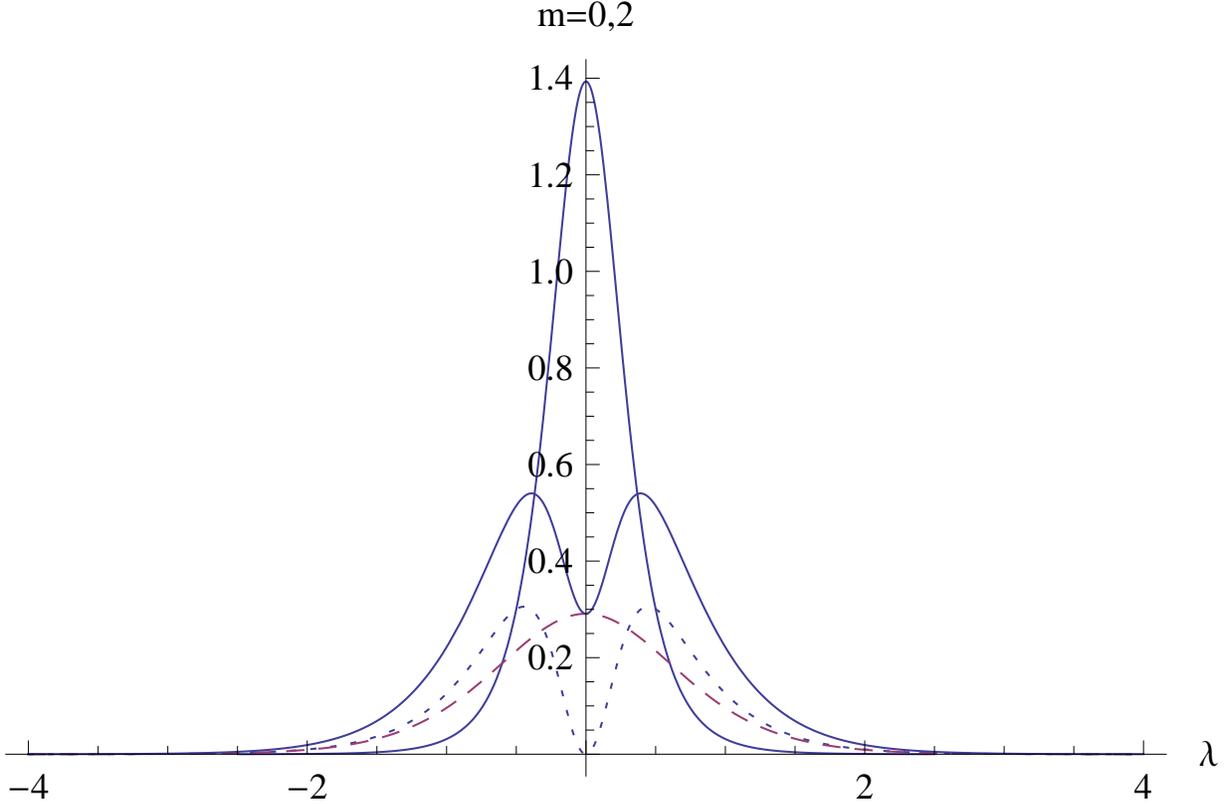}%
\caption{Distribution density of $Q_{x}$ for the ground state $1/\protect\sqrt{2\protect\pi })$ (solid line with highest peak near at $1.4$), and
that for 2nd excited state $\exp {(2i\protect\varphi )}/\protect\sqrt{2\protect\pi })$ which is the sum of $|\protect\alpha (\protect\lambda )|^{2}$
(dotted) and $|\protect\beta (\protect\lambda )|^{2}(\neq |\protect\alpha (\protect\lambda )|^{2})$ (dashed). Neither has a node.}%
\label{figure 1}
\end{figure}

\begin{figure}
\includegraphics[width=0.9\textwidth]{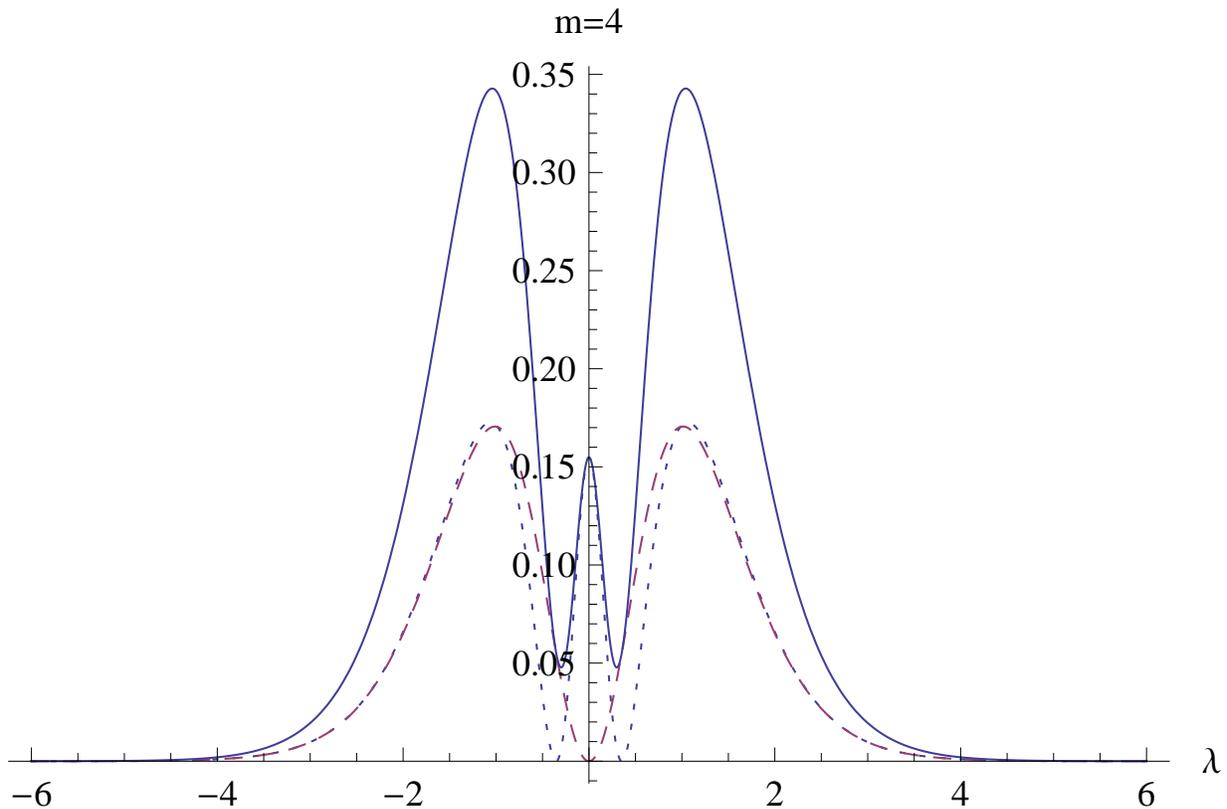}%
\caption{Distribution density of $Q_{x}$ for the 4th excited state $\exp {(4i\protect\varphi )}/\protect\sqrt{2\protect\pi }$, which is the sum of $|\protect\alpha (\protect\lambda )|^{2}$ (dotted) and $|\protect\beta (\protect\lambda )|^{2}$ (dashed) but $|\protect\alpha (\protect\lambda )|^{2}
$ and $|\protect\beta (\protect\lambda )|^{2}$ differ appreciately only near
$\protect\lambda =0$. This density has no node either.}\label{figure 2}
\end{figure}

\begin{figure}
\includegraphics[width=0.9\textwidth]{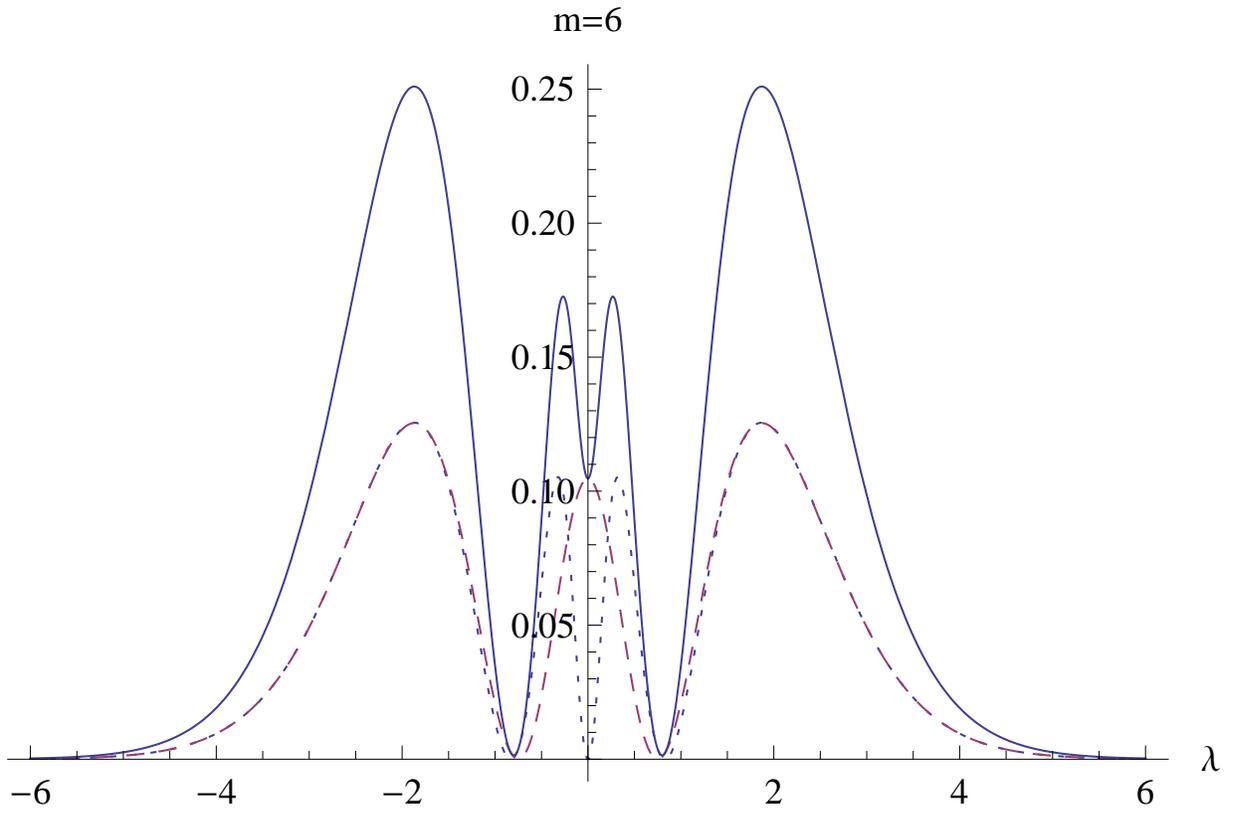}%
\caption{Distribution density of $Q_{x}$ for the 6th excited state $\exp {(6i\protect\varphi )}/\protect\sqrt{2\protect\pi }$, which is the sum of $|\protect\alpha (\protect\lambda )|^{2}$ (dotted) and $|\protect\beta (\protect\lambda )|^{2}$ (dashed) and we see again that $|\protect\alpha (\protect\lambda )|^{2}$ and $|\protect\beta (\protect\lambda )|^{2}$ differ
appreciately only near $\protect\lambda =0$. The density exhibits no node
but two minimum points over interval of finite $\protect\lambda $ almost
reach the zero.}\label{figure 3}
\end{figure}

\begin{figure}
\includegraphics[width=0.9\textwidth]{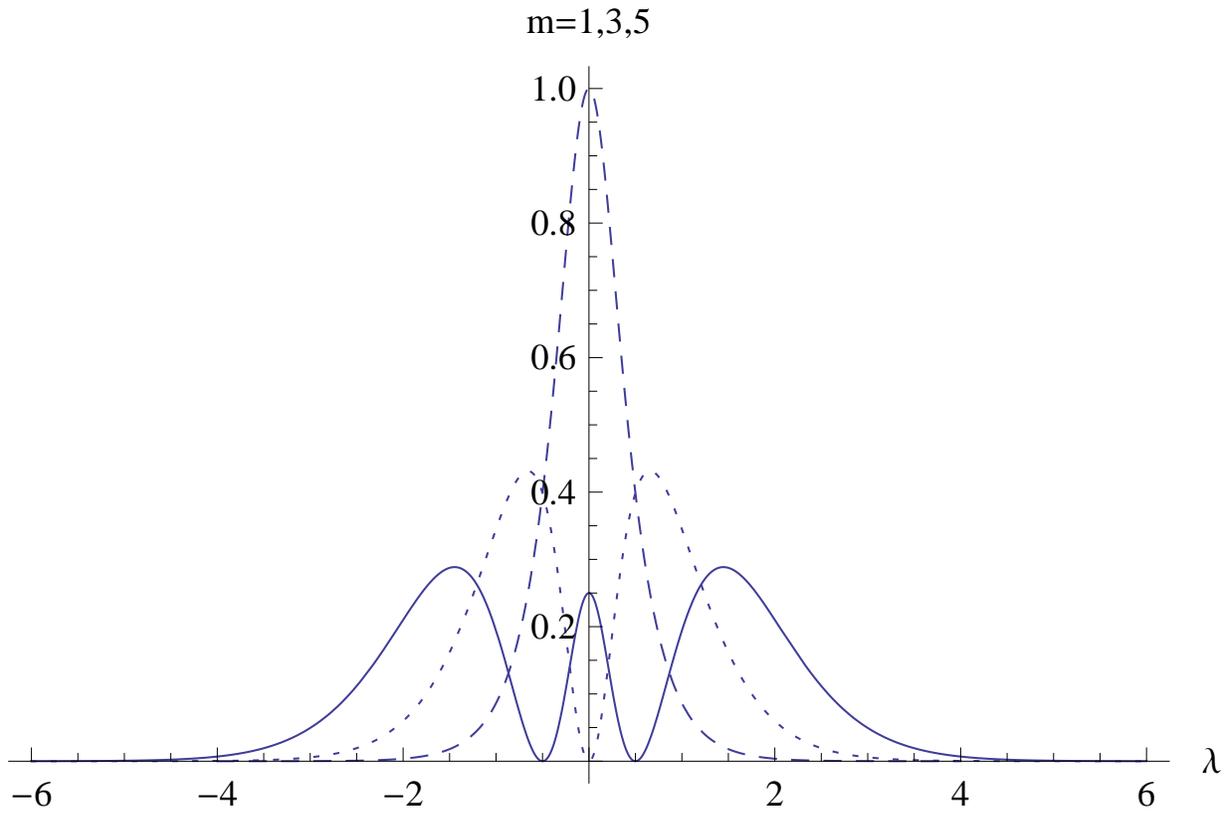}%
\caption{Distribution density of $Q_{x}$ for the 1st (dashed), 3rd (dotted)
and 5th (solid) state $\exp {(im\protect\varphi )}/\protect\sqrt{2\protect\pi })$, $(m=1,3,5)$. These distribution densities and the momentum
distribution densities for the 0th, 1st and 2nd state of one-dimensional
simple harmonic oscillator (not shown), are respectively similar.}\label%
{figure 4}
\end{figure}

\begin{figure}
\includegraphics[width=0.9\textwidth]{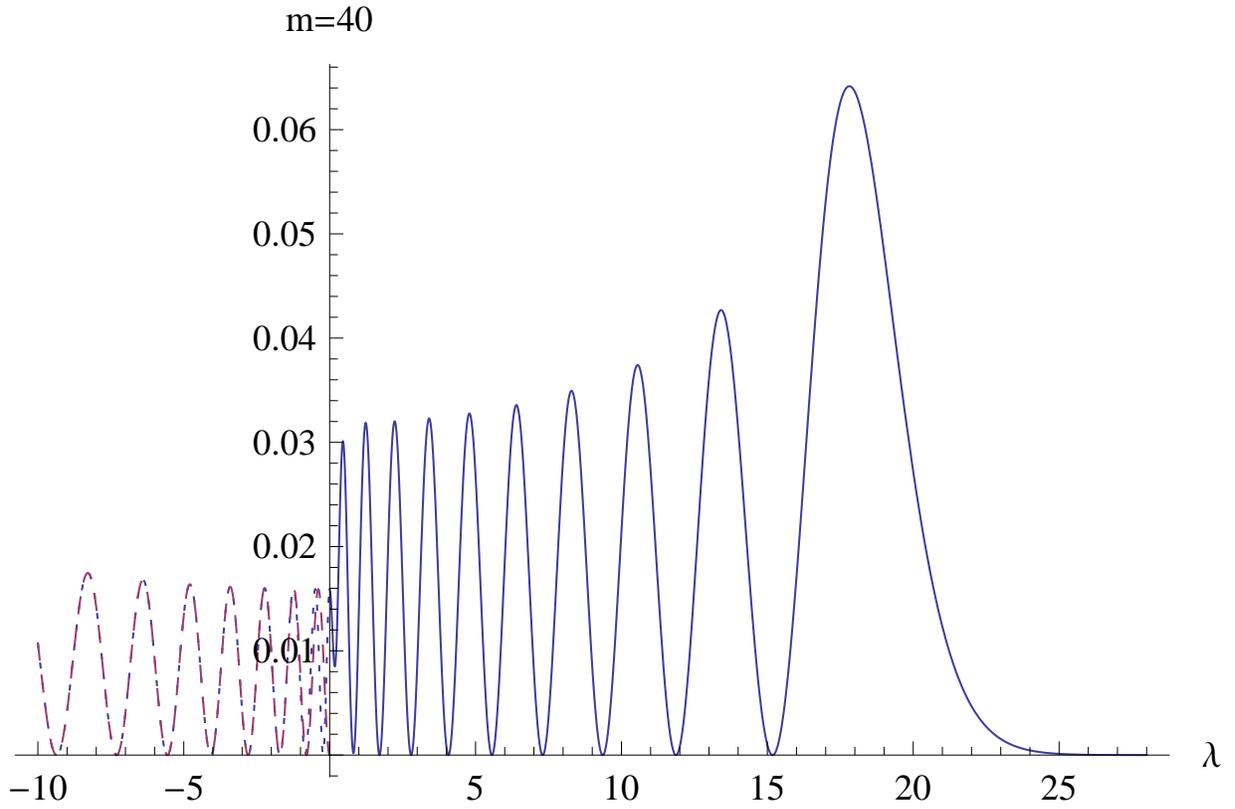}%
\caption{Distribution density of $Q_x$ for the 40th excited state
$\exp{(40i\varphi)}/\sqrt {2\pi}$, which is symmetrical about $\lambda=0$ but
only portion over the postive $\lambda$ is depicted. It is the
sum of $|\alpha(\lambda)|^2$ (dotted) and $|\beta(\lambda)|^2$ (dashed) and
they differ only near point $\lambda=0$; both symmetrical about $\lambda=0$
but half portions over the negative $\lambda $ are plotted. This distribution density
has clearly $21$ peaks and $20$ minima in the interval of finite $|\lambda|<\infty $
but apparently $18$ nodes. To note that two minima near $\lambda=0$ appraoch closer as if
they are a single one. We can infer that in the limit of large $m$
that is even, it is more and more similar to the distribution density for
the $(m/2-1)$th state of one-dimensional simple harmonic oscillator.}\label%
{figure 5}
\end{figure}

\begin{figure}
\includegraphics[width=0.9\textwidth]{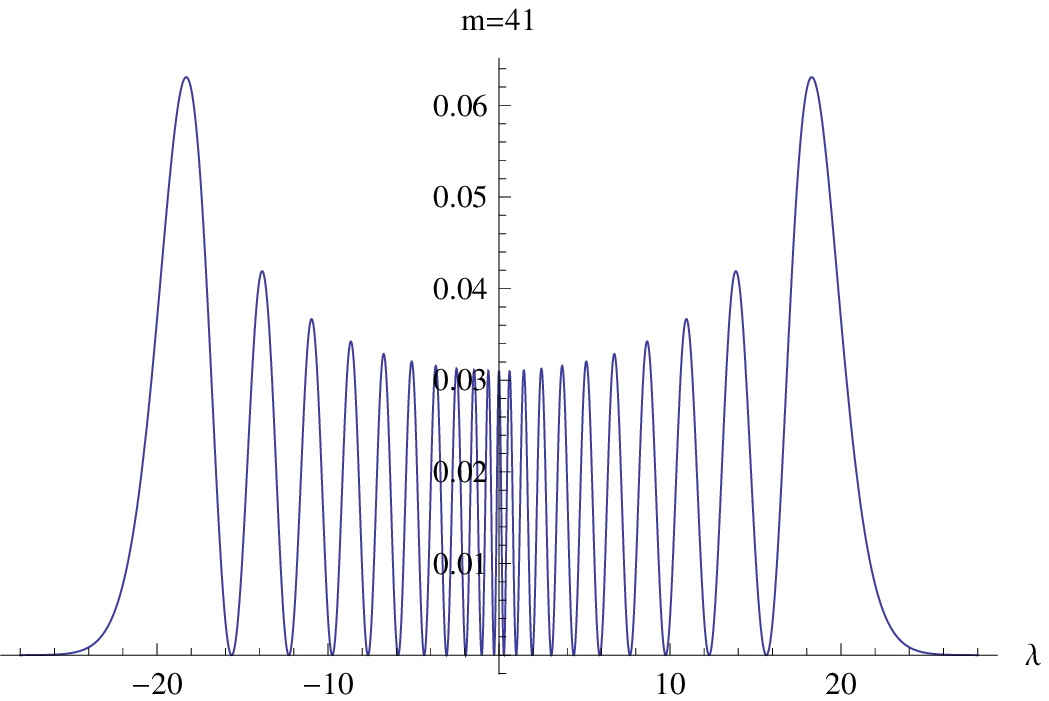}%
\caption{Distribution density of $Q_x$ for the 41th excited state
$\exp{(41i\varphi)}/\sqrt {2\pi})$, which is the sum of two identical part
$|\alpha(\lambda)|^2$ and $|\beta(\lambda)|^2$.
The distribution density and the momentum distribution density for
the $20$th excited state of one-dimensional simple harmonic oscillator (not shown),
are almost the same. So, we can infer that in the limit of large $m$
that is odd, it becomes more and more similar to the distribution density for
the $\{(m-1)/2\}$th state of one-dimensional simple harmonic oscillator.}%
\label{figure 6}
\end{figure}


\section{Conclusions}

The posmom offers a new way to understand the quantum motions, constrained
or not. This study explores the posmom on a circular circle, and identify
that the momentum in it is the geometric momentum that is recently proposed
to properly describe of the momentum for the motions constrained on the
curved surface.\ For construction of the complete basis, we need to resort
to the mutual commutativity between posmom and parity operators, and then
obtain the satisfactory bases each of them is in general four-valued. The
posmometry of the eigenstates of the $z$-axis component of the angular
momentum is worked out, and is found to be similar to the momentum
distributions of stationary states for the one-dimensional simple harmonic
oscillators. Then any states on the circle can thus go through the
posmometry analysis. Once the posmometer is successfully designed and built
up, the ground state of the planar rotation of some molecules, which can be
easily prepared, can be visualized via the distribution of density of the
posmom.

The present exploration riches not only our appreciation of the quantum
dynamical behavior, but also our understanding of the fundamental aspect of
the quantum mechanics.

\begin{acknowledgments}
This work is financially supported by National Natural Science Foundation of
China under Grant No. 11175063.
\end{acknowledgments}

\end{subequations}

\end{document}